\documentstyle[11pt]{article}

 \csname
@addtoreset\endcsname{equation}{section}

\def\a{\alpha}

\def\c{\gamma}
\def\C{\Gamma}
\def\d{\delta}

\def\e{\epsilon}
\def\ve{\varepsilon}
\def\f{\phi}

\def\vf{\varphi}
\def\k{\kappa}
\def\l{\lambda}

\def\m{\mu}
\def\n{\nu}
\def\r{\rho}
\def\s{\sigma}

\def\o{\omega}
\def\pb{{\bar\psi}}
\def\eb{{\bar\varepsilon}}
\def\lb{{\bar\lambda}}
\def\lra{\leftrightarrow}

\def\g{1+\e |\phi|^2}

\def\fs{|\phi|^2}
\def\ra{\rightarrow}
\def\lra{\leftrightarrow}
\def\qq{\quad\quad}

\def\del{\partial}
\let\la=\label
\let\bm=\bibitem
\def\nn{\nonumber}
\newcommand{\eq}[1]{(\ref{#1})}

\newcommand{\w}[1]{\\[0.#1cm]}

\def\be{\begin{equation}}
\def\ee{\end{equation}}
\def\bea{\begin{eqnarray}}
\def\eea{\end{eqnarray}}
\def\ba{\begin{array}}
\def\ea{\end{array}}

\def\ft#1#2{{\textstyle{{\scriptstyle #1}
\over {\scriptstyle #2}}}}

\def\ed{\end{document}}

\newcommand{\hoch}[1]{$\, ^{#1}$}

\newcommand{\tamphys}{\it\small Center for Theoretical Physics, Texas
A\&M University, College Station, TX 77843, USA}

\newcommand{\karlstad}{\it\small Physics Department, Karlstad
University, S-651 88 Karlstad, Sweden}

\newcommand{\auth}{\large N.S. Deger \hoch{1}, A. Kaya \hoch{1}, E.
Sezgin \hoch{1\dagger} and P. Sundell \hoch{2}}

\begin{document}

\hfill{CTP-TAMU-20/99}

\hfill{hep-th/9908089}

\hfill{\today}

\vspace{20pt}

\begin{center}

{\Large \bf Matter Coupled AdS$_3$ Supergravities \\
\smallskip and Their Black Strings }

\vspace{30pt}


\auth

\vspace{15pt}

\begin{itemize}

\item[$^1$] \tamphys

\item[$^2$] \karlstad

\end{itemize}

\vspace{30pt}

{\bf Abstract}

\end{center}

We couple $n$ copies of $N=(2,0)$ scalar multiplets to a gauged
$N=(2,0)$ supergravity in $2+1$ dimensions which admits $AdS_3$ as a
vacuum. The scalar fields are charged under the gauged $R$-symmetry
group $U(1)$ and parametrize certain Kahler manifolds with compact or
non-compact isometries. The radii of these manifolds are quantized in
the compact case, but arbitrary otherwise. In the compact case, we find
half-supersymmetry preserving and asymptotically Minkowskian black
string solutions. For a particular value of the scalar manifold radius,
the solution coincides with that of Horne and Horowitz found in the
context of a string theory in $2+1$ dimensions. In the non-compact
case, we find half-supersymmetry preserving and asymptotically $AdS_3$
string solutions which have naked singularities. We also obtain two
distinct $AdS_3$ supergravities coupled to $n$ copies of $N=(1,0)$
scalar multiplets either by the truncation of the $(2,0)$ model or by a
direct construction.

{\vfill\leftline{}\vfill \vskip 10pt \footnoterule {\footnotesize
\hoch{\dagger} Research supported in part by NSF Grant PHY-9722090.
\vskip -12pt}

\pagebreak

\setcounter{page}{1}


\section{Introduction}


Important advances have been made in our understanding of $M$-theory in
$AdS$ space over the last few years \cite{malda,gubser,witten}. In particular,
evidence has been accumulated for a remarkable relation between certain 
gauged  supergravities which admit $AdS$ space as vacua and  appropriate
conformal 
field theories defined on the boundaries of these $AdS$ spaces. An especially 
manageable  example of this phenomenon arises in the context of $AdS_3/CFT_2$ 
correspondence. While interesting work has been done on the CFT aspects of this 
problem,  a great deal remains to be done on the supergravity side. With this 
motivation  in mind,  and in view of their relative simplicity, in this
paper we  
study the  structure of matter coupled$AdS_3$ supergravity theories with
$N=(2,0)$ 
and $N=(1,0)$  supersymmetry and their string solutions.  

Since the $AdS_3$ group is reducible, namely, $SO(2,2) \approx
SO(2,1)_I\times SO(2,1)_{II}$, the supersymmetry parameters could come
in $p$ copies of $SO(2,1)_I$ and $q$ copies of $SO(2,1)_{II}$, thus
describing $(p,q)$ supersymmetry. The pure $AdS_3$ supergravity with
$(p,q)$ supersymmetry was constructed long ago and it is a Chern-Simons
gauge theory based on the $AdS_3$ supergroup $OSp(p|1) \oplus OSp(q|1)$
\cite{at}. Later, the coupling of the $N=(2,0)$ case to $n$ copies of
scalar multiplets which parametrize a Kahler manifold was constructed
\cite{it}. In this model, the scalars are neutral under the 
$R$-symmetry group $U(1)$. Supersymmetric solutions of this model have been
studied and in particular, it has been shown \cite{it} that the pure
$AdS_3$ supergravity sector of the theory  admits the BTZ black hole
\cite{btz} as a supersymmetric solution. The model has features unlike
the familiar gauged supergravity theories. In fact, a higher
dimensional origin of the model, whether it is field theoretic one or
string/M-theoretic, is apparently not known.

There must exist, however, a class of (matter coupled) $AdS_3$
supergravity theories which describe various $AdS_3$ compactifications
of string/M-theories.  In particular, the $AdS_3\times S^3\times K_3$
compactification of Type IIB string has been a subject of number of
studies recently \cite{jmas,dkss,db}.  The full spectrum of this
compactification is known, and the massless sector is expected to be
described by a matter coupled $N=(4,4), AdS_3$ supergravity with the
gauge group $SO(3)_L\times SO(3)_R$.  There exist other
supersymmetric compactifications of supergravities in diverse dimensions 
down to $AdS_3$, and in all these cases, we expect to find the
gauged versions of the matter coupled Poincar\'e supergravities in $2+1$ 
dimensions constructed long ago by Marcus and Schwarz \cite{ms}, with the 
matter sector consisting of scalar multiplets with an underlying
$SO(8,n)/SO(n)\times SO(8)$ (for certain values of $n$ implied by string
theory) or $E_8/SO(16)$ structure, and their lower supersymmetric
truncations.

Ultimately we would like to construct all the $AdS$ supergravities
mentioned above in a unified framework, and to study their connection with
the boundary conformal field theories.  In this paper we take the first
step in this direction. In particular, we construct an $N=(2,0), AdS_3$
supergravity coupled to $n$-complex dimensional Kahler manifold, and its
$N=(1,0)$ truncation.  This paper is devoted to understanding of the
supergravity aspects of these models and their string solitons.
We expect that there exist compactifications of  string/M-theory giving 
rise  to the supergravity theories studied here as their low energy limits.

As mentioned above there already exists a matter coupled
$N=(2,0)$ $AdS_3$ supergravity constructed sometime ago by Izguierdo
and Townsend \cite{it}. However, this model differs from ours in a
significant way, namely its scalar fields are neutral with respect to
the $R$-symmetry group $U(1)$ unlike in our model. Consequently, the
model of \cite{it} does not have a potential while our model leads to a
rather elaborate potential. In fact, some of the properties exhibited
by our model are quite similar to those which arise in gauged $N=1$
supergravity coupled to a restricted kind of Kahler sigma model in
$D=4$ \cite{wb}. For example, the sigma model manifold arising can be
either compact or non-compact. The gravitational coupling constant, the
radius of $AdS_3$ and the radius of the sigma model manifold are not
related to each other by supersymmetry, unlike the gauged
supergravities with higher supersymmetries. Moreover, the radius of the
sigma model manifold, when compact, is quantized in units of the
gravitational coupling constant.

We also obtain a $N=(1,0)$ supersymmetric version of the results
mentioned above by a consistent truncation of the $N=(2,0)$ model. We
show that there exists a one parameter extension of this $N=(1,0)$
theory for the coupling of single scalar multiplet.

In this paper we also present string solutions of our models
which preserve half supersymmetry, both for compact and non-compact
sigma models. Interestingly enough, for the compact case there is a
particular value of the radius such that the solution reduces to that
of Horne and Horowitz \cite{hh} found in the context of low energy
limit of a string theory in $2+1$ dimensions. The solutions exhibit an 
event horizon and are  asymptotically Minkowskian for the compact sigma 
model.  The solutions for the non-compact sigma model, on the other hand, 
have naked singularities and are asymptotically $AdS_3$. 

The plan of this paper is as follows. In Sec. 2, we describe the
$N=(2,0)$ $AdS_3$ supergravity coupled to an $n$-complex dimensional
Kahler sigma model. In Sec. 3, we specialize to the case of $n=1$,
namely $CP^1$ and $CH^1$. The black string solutions and their
properties will be discussed in Sec. 4. The $N=(1,0)$ supersymmetric matter 
coupled $AdS_3$ supergravity is presented in Sec. 5. Our results and a number 
of open problems raised by them are discussed
in Sec. 6.


\section{N=(2,0)\ AdS$_3$ Supergravity Coupled to $n$ Scalar Multiplets}


The $N=(2,0)\ AdS_3$ supergravity multiplet consists of a graviton
$e_\m{}^a$, two Majorana gravitini $ \psi_\m$ (with the $SO(2)$ spinor
index suppressed) and an $SO(2)$ gauge field $A_\m$. The $n$ copies of
the $N=(2,0)$ scalar multiplet, on the other hand, consists of $2n$
real scalar fields $\f^\a (\a=1,...,2n)$ and $2n$ Majorana fermions
$\l^r\ (r=1,...,n\ {\rm and\ the }\ SO(2)\ {\rm spinor\ indices\ are\
suppressed})$.

For simplicity, we will take the sigma model manifold $M$ to be a coset
space of the form $G/H\times SO(2)$ where $G$ can be compact or
non-compact and $H\times SO(2)$ is the maximal compact subgroup of $G$,
where $SO(2)$ is the $R$-symmetry group. For concreteness, we shall
consider

\be
M_+= {SO(n+2)\over SO(n)\times SO(2)}\ ,
\quad\quad
M_-= {SO(n,2)\over SO(n)\times SO(2)}\ .
\la{pm}
\ee

Our results can be readily translated to the case of $G/H\times U(1)$ with
$G=SU(n+1)$ of $SU(n,1)$ and $H=SU(n)$. \\

Let $(L_I{}^i, L_I{}^r)$ where $I=1,...,n+2,\  i=1,2,\  r=1,..,n$ form a
representative of the coset $M_\pm$. It follows that

\bea
&& L_I{}^i L^{Ij} = \pm \d^{ij}\ , \quad\quad L_I{}^r L^{Is}=\d^{rs}\ ,
\quad\quad L_I{}^i L^{Ir} = 0\ ,
\la{cons}\w2
&& \pm L_I{}^iL^{Ji} + L_I{}^r L^{Jr} = \d_I^J\ ,
\nn
\eea

where $\pm$ correspond to the scalar manifolds $M_{\pm}$. The $SO(n)$,
$SO(2)$ and $SO(n+2)$ vector indices are raised and lowered with the
Kronecker deltas and the $SO(n,2)$ vector indices with the metric
$\eta_{IJ}= {\rm diag}(++...+--)$.

The $SO(2)$ gauged pull-back of the Maurer-Cartan form on $M_\pm$ can be
decomposed into the $SO(n)\times SO(2)$ connections $Q_\m^{rs}$ and
$Q_\m^{ij}$, and the nonlinear covariant derivative $P_\m^{ir}$
as follows:

\be
P_\m^{ir} = \left(L^{-1} D_\m L\right)^{ir}\ ,
\ \ \ \
Q_\m^{ij} = \left(L^{-1} D_\m L\right)^{ij}\ ,
\ \ \ \
Q_\m^{rs} =  \left(L^{-1} D_\m L\right)^{rs}\
\la{mc}
\ee

where $Q_\m^{ij} \equiv Q_\m \e^{ij}$ and the $SO(2)$ covariant
derivative is defined as

\be
D_\m L= \left( \del_\m +\ft12 A_\m^{ij} T_{ij}\right)L \ ,
\quad\quad
A_\m^{ij}\equiv A_\m \e^{ij}\ .
\ee

The anti-hermitian $SO(2)$ generator $T_{ij}$ occurring in this
definition is realized in terms of an $\footnotesize{(n+2)\times
(n+2)}$ matrix, which can be chosen as $(T_{ij})_I{}^J= (\,\pm\,
\d_{Ii}\,\d_{j}^J - i \leftrightarrow j\,) $.

In coupling $M_\pm$ to supergravity, we will also need the introduction
of the ``boosted matrix elements'' defined as

\bea
\e_{ij}\e^{kl} C &=& (L^{-1}T_{ij}L)^{kl}\ ,
\nn\w2\
\e_{ij}C^{rs} &=& (L^{-1}T_{ij}L)^{rs}\ ,
\nn\w2
\e_{ij}S^{kr} &=& (L^{-1}T_{ij}L)^{kr}\ .
\la{cf}
\eea

From these definitions and \eq{mc} it follows that

\bea
\del_{[\m} Q_{\n]}&=&\ft12\e_{ij} P_\m^{ir} P_\n^{jr}+\ft12 F_{\m\n} C\ ,
\nn\w2
D_{[\m} P_{\n]}^{ir}&=& \ft12 F_{\m\n} S^{ir}\ ,
\nn\w2
\del_\m C &=& \e_{ij} P_\m^{ir} S^{jr}\ ,
\la{lemmas}\w2
D_\m C^{rs} &=& \pm 2 P_\m^{i[r} S^{s]i} \ ,
\nn\w2
D_\m S^{ir} &=& \pm \e_{ij} P_\m^{jr} C + C^{rs} P_\m^{is}\ ,
\nn
\eea
\smallskip

where the covariant derivatives are defined as

\bea
D_\m S^{ir} &:=& \del_\m S^{ir}\pm Q_\m^{ij}S^{jr} + Q_\m^{rs} S^{is}\ ,
\nn\w2
D_\m C^{rs} &:=& \del_\m C^{rs}+ Q_\m^{rs} S^{is}\ .
\eea

Recall that $\pm$ correspond to the scalar manifolds $M_\pm$ specified
in \eq{pm}.

Using the formulae given above and applying the standard Noether
procedure we get the following matter coupled gauged supergravity
Lagrangian up to quartic order fermion terms
\footnote{\footnotesize {Our conventions are as follows:
$\eta_{ab}=(-++), \ \eb=\ve^\dagger i\c_0$, $\c^\m C$ and $\c^{\m\n}C$
are symmetric, the $SO(2)$ charge conjugation matrix is unity,
\,$\C^i$\, is symmetric and $\{\C^i,\C^j\}=2\d^{ij}$. A
convenient
representation is $\C_1=\s_1$, $\C_2=\s_3$. We define $\C_3=\C_1\C_2$.
Note that $(\C^3)^2=-1$.}}

\bea
e^{-1}{\cal L} &=&  {1\over 4} R + {1\over 2} \e^{\m\n\r}\pb_\m D_\n \psi_\r
-{1\over 16}{e^{-1}\over ma^4}\, \e^{\m\n\r} A_\m \del_\n A_\r
-{1\over 4a^2} P_\m^{ir} P^\m_{ir}
\nn\w2
&& +{1\over 2}\,\lb_r\c^\m D_\m \l^r
   + {1\over 2a}\, \lb_r \c^\m\c^\n \C_i\psi_\m P_\n^{ir}
   -{m\over 2}\, \pb_\m\c^{\m\n}\psi_\n C^2
\nn\w2
&&  -2 ma\,\pb_\m \c^\m\C_i\C^3\l_r C S^{ir}
    -{1\over 2} m (1\pm 4 a^2)\, \lb^r\l_r C^2
\nn\w2
&& + {2 ma^2}\,\lb_r\C^3\l_s C^{rs} C
+ {2 ma^2}\,\lb_r\C_i\C_j\l_s S^{ir}S^{js}
\nn\w2
&& +{2 m^2} C^2(C^2 -2 a^2 S^{ir}S_{ir})\ ,
\la{a}
\eea

which has the local $N=2$ supersymmetry

\bea
\d e_\m{}^a &=& -\eb\c^a\psi_\m\ ,
\nn\w2
\d \psi_\m &=& D_\m \ve +m \c_\m C^2 \ve\ ,
\nn\w2
\d A_\m &=& 4 m a^2\, (\eb \C^3\psi_\m)\, C
- 4 ma^3\,(\lb_r\c_\m\C^i\e)\, S^{ir}\ ,
\nn\w2
L_i{}^I \d L_I{}^r &=& a\, \eb\, \C_i \l^r\ ,
\nn\w2
\d\l^r &=& \left( -{1\over 2a} \c^\m P_\m^{ir}
+ 2 ma \C^3 C S^{ir}\right) \C_i\ve\ .
\la{s}
\eea

We have set the gravitational coupling constant $\k$ equal to one, but
it can easily be introduced by dimensional analysis. The constant $a$
is the characteristic curvature of $M_\pm$ (e.g. $2a$ is the inverse
radius in the case of $M_+=S^2$) and the constant $m$ is the $AdS_3$
cosmological constant. The $U(1)$ gauge coupling constant has been
absorbed into the definition of $A_\m$. We emphasize that, unlike in a
typical anti de Sitter supergravity coupled to matter, here the
constants $\k,a,m$ are not related to each other for non-compact scalar
manifolds, while $a$ will be quantized in terms of $\k$ in the compact
case, as we shall see later.

The covariant derivatives are defined as

\bea
D_\m \ve &=& \left(\del_\m +\ft14 \o_\m{}^{ab} \c_{ab}
-{1\over 2 a^2}\,Q_\m \C^3\right)\ve\ ,
\nn\w2
D_\m \l^r &=&\left(\del_\m +\ft14 \o_\m{}^{ab} \c_{ab}
+{1\over 2a^2} Q_\m \C_3 \right)\l^r +Q_\m{}^{rs}\l^s\ .
\la{d}
\eea

The coefficients in front of the composite connection $Q_\m$ has been
determined by supersymmetry.

Having defined the above covariant derivative, we can now see more
clearly why the $C$- and $S$-functions arise in the model. Firstly, the
supersymmetric variation of the gravitino kinetic term involves the
commutator

\be
[D_\m,D_\n]\ve = \ft14 R_{\m\n ab} \c^{ab} \ve -{1\over 2a^2}
\left(\e_{ij} P_\m^{ir} P_\n^{jr}+ F_{\m\n} C\right) \C^3\ve\ .
\la{com}
\ee

This is where we see first the occurrence of the function $C$. The
$C$-dependent term arising here is cancelled by the variation of the
Chern-Simons term. The rest of the Noether procedure eventually
involves the differentiation of the function $C$ which leads to the
function $S^{ir}$, and its differentiation leads to the function
$C^{rs}$.

It is straightforward to adapt all the formulae given above in terms of
$n$ complex scalars $\f^\a (\a=1,...,n)$ and $n$ Dirac spinors $\l^r$
and a Dirac gravitino $\psi_\m$, with the dreibein $e_\m^a$ and vector
field $A_\m$, of course, remaining real. Typical sigma model manifolds
arising in this way are the compact $CP^n$ and non-compact $CH^n$, if
we are only concerned about the local aspects of the symmetries
involved. Insisting that the model is globally well defined, some
restrictions will arise on the scalar manifold geometry, as was shown
long ago by Witten and Bagger in the context of $N=1$ supersymmetric
models coupled to supergravity in $D=4$. These restrictions typically
occur when the scalar manifold is compact. Indeed, as in \cite{wb},
here too the compact scalar manifold turns out to be a Hodge manifold,
which is a certain type of Kahler manifold. An important consequence of
this is that the radius of the scalar manifold gets quantized in units
of the Planck length. This phenomenon is explained in detail in
\cite{wb} and therefore will not be repeated here. However, we shall
get back to the specifics of the quantization condition in the next
section where we consider the black string solutions of the model when
the scalar manifold is in effect taken to be $S^2$. In the case of
$H^2$ the quantization condition does not arise. \\

Let us consider various limits of this model. Firstly, by rescaling
$A_\m \ra a^2 A_\m$ and the matter scalar fields such that $P_\m^{ir}
\ra aP_\m^{ir}$ and then sending the inverse sigma model radius $a\ra
0$ such that $C\ra 1$, $S^{ir} \ra 0$, one obtains $N=(2,0)$, $AdS_3$
supergravity with cosmological constant $-2m^2$ coupled to an $R^{2n}$
sigma model. The pure $N=(2,0)$, $AdS_3$ supergravity \cite{at} can
then be obtained by setting all the matter fields equal to zero. To
obtain the Poincar\'e limit of the theory \cite{ms,dw,ht} , on the
other hand, we start with the Lagrangian \eq{a}, rescale $A_\m \ra m
A_\m$ and then let $m \ra 0$. Once the Poincar\'e limit is taken, the
supergravity fields can further be decoupled by setting the gravitini
equal to zero and taking the metric to be flat Minkowskian. This yields
an $N=(2,0)$ supersymmetric sigma model. A rigid $AdS_3$ supersymmetric
limit does not seem to be possible in this model.


\section{The Cases of $S^2$ and $H^2$}


The variables of the model presented above can easily be complexified
so that the scalar manifold becomes generically
$CP^n=SU(n+1)/SU(n)\times U(1)$ or $CH^n=SU(n,1)/SU(n)\times U(1)$. In
search of a string solution, it is convenient to set equal to zero all
but one of the $n$-complex scalar fields in such a way that the model
is consistently truncated to an $S^2=SU(2)/U(1)$ or $H^2=SU(1,1)/U(1)$
sigma model coupled to the $AdS_3$ supergravity with $N=(2,0)$
supersymmetry. The $R^2$ geometry can easily be accounted for as the
infinite radius limit of $H^2$.

The coset representative $L$ for $S^2$ and $H^2$ can be parametrized as

\be
L={1\over \sqrt{1+ \e\fs}}\left(\matrix{1 &
\f \cr -\e\f^\dagger & 1\cr}\right)\ ,
\quad\quad
\e =\cases{
+1 \ \ \ {\rm for}\ \ \ S^2\cr
-1 \ \ \ {\rm for}\ \ \ H^2\cr}
\la{cr}
\ee
\\

Defining

\bea
SU(2): & S=S^1+iS^2\ ,& P_\m= -iP_\m^1+ P_\m^2\ ,\nn \\
SU(1,1):& \ \ S=-iS^1+S^2\ ,& P_\m= P_\m^1+ i P_\m^2\ ,
\la{cs2}
\eea

the key relations \eq{lemmas} take the form

\bea
\del_\m C &=& -\e\, (P_\m^* S + P_\m S^*)/2\ ,
\nn\w2
D_\m S &=& P_\m C\ .
\la{cn}
\eea

The functions $C$ and $S$ are computed from the definitions \eq{cf},
which, for the cases at hand, are

\be
L^{-1}T_3L=\ft12 (S_1+iS_2)(T_1-iT_2) + \ft12 (S_1-iS_2)(T_1+iT_2) + CT_3\ ,
\la{dc}
\ee

where the generators of $SU(2)$ and $SU(1,1)$ algebras are

\be
[T_1,T_2]=-T_3,\qq [T_2,T_3]=-\e T_1\ ,\qq  [T_3,T_1]=-\e T_2 \ .
\ee

Representing $(T_1,T_2,T_3)$ by $(-i\s_1,i\s_2, -i\s_3)/2$ for $SU(2)$
and $(\s_1,-\s_2,i\s_3)/2$ for $SU(1,1)$, we obtain from \eq{dc}

\bea
C&=& {1- \e\fs \over \g}\ ,
\nn\w2
S &=& {2\f \over\g}\ .
\la{cs1}
\eea

Similarly, the nonlinear covariant derivative $P_\m$ and the $SO(2)$
connection $Q_\m$ are computed from the definitions \eq{mc}, which for
the cases we are considering, take the form

\be
L^{-1}(\del_\m +A_\m T_3) L = \ft12 (P^1_\m+i P^2_\m)(T_1-iT_2) +
\ft12 (P^1_\m-i P^2_\m)(T_1+iT_2) + Q_\m T_3\ ,
\la{pq2}
\ee

from which, recalling the definitions \eq{cs2}, it follows that

\bea
P_\m &=& {2 D_\m\f\over \g}
\nn\w2
     &=& {2 \del_\m\f\over \g}  -i\e  A_\m S\ ,
\nn\w2
Q_\m &=& {i \f \stackrel{\lra}{D_\m}\f^*  \over \g}\,+A_\m
\nn\w2
&=&{i \f \stackrel{\lra}{\del_\m}\f^*  \over \g}\,+A_\m C\ ,
\eea

where

\be
D_\m \f = \left(\del_\m -i\e A_\m \right)\f\ .
\ee

In describing the sigma model manifolds, we have used a particular
coordinate system. We have to ensure that this description makes sense
globally. In fact, the coordinates $\f$ are the stereographic
projection of $S^2$ or $H^2$ onto the complex plane. In the case of
$H^2$ this is a globally well defined coordinate system to cover the
manifold. But in the case of $S^2$, as is well known, one needs two
patches in order to avoid the singularities at the north and south
poles.  Following the standard procedure, we cover the upper
hemisphere with coordinate $\f$ and the lower one by $ 1/\f$. We must
then check that the action is well defined in the overlap region. To
achieve this, we also need to transform the gauge field as $A_\m \ra
-A_\m$. Under the combined transformation

\be
\f \ra {1\over\f}\ ,\qq A_\m \ra -A_\m\ ,
\ee

the  quantities $C,S,P_\m$ and $Q_\m$ transform as

\bea
C &\ra& -C \ ,
\nn\w2
S &\ra& \left({\f^* \over \f}\right)\, S\ ,
\la{k}\w2
 P_\m &\ra&  - \left({\f^* \over \f}\right) P_\m\ ,
\nn\w2
Q_\m &\ra& Q_\m +i\del _\m\,(\,{\rm ln}\,\f-{\rm ln}\,\f^*\,)\ .
\nn
\eea

For these transformations to leave the action invariant, we must also
transform the fermionic fields. Noting that these terms are given by

\be
\ft12 \e^{\m\n\r} {\pb}_\m \left(\nabla_\n -{i\k^2\over 2a^2}
Q_\n\right)\psi_\r + \ft12 \lb\c^\m \left(\nabla_\m-{i\k^2\over 2a^2}
Q_\m\right)\l\ ,
\ee

where $\nabla_\m$  is the Lorentz covariant derivative, we find that the
appropriate transformation rules for the fermions are

\bea
\psi_\m & \ra & {\rm exp} \left[-{\k^2\over 2a^2} (\,{\rm ln}\,\f
- {\rm ln}\,\f^*\,)\right]\ \psi_\m\ ,
\nn\w2
\l & \ra & {\rm exp} \left[ -{\k^2\over 2a^2} (\,{\rm ln}\,\f
- {\rm ln}\,\f^*\,)\right]\ \l\ ,
\eea

where we have re-introduced the gravitational coupling constant $\k$.
For these transformations to be single valued, we need to impose, \'a
la Witten and Bagger \cite{wb}, the quantization condition

\be
{\k^2\over a^2}\,=\,n \ ,
\ee
\\

where $n$ is an integer.


\section{The Black String Solution}


We shall now seek string solutions of the model described in the
previous section. To this end, let us note the bosonic part of the
Lagrangian

\be
e^{-1}{\cal L} = {1\over 4} R
-{1\over 16}{e^{-1}\over ma^4}\, \e^{\m\n\r} A_\m \del_\n A_\r
- { |D_\m\f|^2 \over a^2(1+\e |\f|^2)^2} -V(\f)\ ,
\la{ba1}
\ee

where the potential is given by

\be
V(\f)= 4m^2a^2C^2\left( |S|^2-{1\over 2a^2}C^2\right)\ ,
\la{a2}
\ee
\smallskip

and $C$ and $S$ are defined in \eq{cs1}. Note that $\e |S|^2=1-C^2$. The
resulting bosonic field equations are

\bea
&& R_{\m\n} = {1\over a^2} \,P_{(\m} P_{\n)}^*
+ 4 V g_{\m\n} \ ,
\la{e1}\w2
&& \sqrt{-g} \e_{\m\n\r}  F^{\n\r} =- 4i m a^2\,
( P_\m S^* - P^*_\m S)\ ,
\la{e2}\w2
&& {1\over \sqrt{-g}} \del_\m \left(\sqrt{-g} g^{\m\n} P_\n\right)
 -i\e Q_\m P^\m= 2a^2 \left(1+\e |\f|^2\right)  {\del V\over \del\phi^*}\,
\la{e3}
\eea

where

\be
{\del V\over \del\phi} = -{8\e m^2\over 1+\e|\f|^2}
\left( a^2 |S|^2 -(1+\e a^2)
C^2\right) C S^* \ .
\ee

We shall also need the supersymmetry transformation rules. Let us first
define

\be
{\hat\ve} := {1\over 2} (1-i\C_3)\ve \ , \qq
{\hat\psi_\m} := {1\over 2}(1-i\C_3)\psi_\m\ ,\qq
 {\hat\l} := {1\over 2}(i)^{(1+\e)/2} (1+i\C_3)\l\ .
\ee

Dropping the hat for notational convenience, the transformation rules
\eq{s}, applied to the case at hand, take the form

\bea
\d \psi_\m &=& D_\m \ve +m \c_\m C^2 \ve\ ,
\nn\w2
\d\l &=&\left(- {1\over 2a} \c^\m P_\m + 2\e m a \, CS \right) \C^-\ve\ ,
\la{s2}
\eea

where $\C^{\pm} = (\C^1\pm i\C^2)/2$.

Before presenting the black string solutions,it is worthwhile to note
that the theory admits various maximally symmetric vacua. For the case
of $S^2$,the potential \eq{a2} has minimum at $\f=0$ corresponding to a
supersymmetric $AdS_3$ vacuum, a valley of minima at $\f=e^{i\theta}$
corresponding to a supersymmetric $2+1$ dimensional Minkowski vacuum
and two valleys of maxima at $\f_\pm =(1\mp \l/1\pm \l)^{1/2}
e^{i\theta}$, where $\l=a/\sqrt{2a^2+1}$, corresponding to
non-supersymmetric de Sitter vacua. Here $\theta$ is an arbitrary real
scalar field. For the case of $H^2$, we have the following extrema: (i)
For $a^2 \le 1/2$, there is a maximum at $\f=0$ which is a supersymmetric
$AdS_3$ vacuum, (ii) for $1/2 < a^2 <1$, there are two valleys of
minima at $\f_\pm= (\pm\l +1 /\pm\l-1)^{1/2}e^{i\theta}$ where
$\l=a/\sqrt{2a^2-1}$ which are non-supersymmetric $AdS_3$ vacua, (iii)
for $a^2 \ge 1$ there is a minimum at $\f=0$ which is a supersymmetric
$AdS_3$ vacuum. The case (ii) similar in nature to a situation
encountered in finding the extrema of the gauged $D=7$ supergravity theory
\cite{mtv}.

Let us now consider the following ansatz for the metric

\be
ds^2 = e^{2A} (-dt^2+dx^2) + e^{2B} dr^2\ ,
\la{m}
\ee

where $A,B$ are functions of the transverse coordinate $r$ only.
Next, we set

\be
\f= |\f|\ , \quad A_\m =0\ , \quad \psi_\m=0\ ,\quad \l =0\ .
\ee
\\
Then, the supersymmetry condition $\d\l=0$ implies that

\be
e^{-B} \f'= 4ma^2 C\f\ ,
\la{susy}
\ee

where the prime indicates differentiation with respect to $r$, provided
that we also impose the condition
\be
\c^1\ve=-\e \ve \ ,
\ee

which means that we are seeking half-supersymmetry preserving
solution. The choice of minus sign is merely for convenience.

The supersymmetry conditions $\d\psi_0=0$ and $\d\psi_2=0$ are
satisfied provided that

\be
A' = 2\e m C^2 e^B\ .
\la{ap}
\ee

 The remaining condition $\d\psi_1=0$ determines the $r$-dependence of
the spinor $\ve$ to be

\be
\ve = S^{1/4a^2}\,(1-\e \c_1)\ve_0\ ,
\ee

where $\ve_0$ is an arbitrary constant spinor. Next, we use \eq{ap} in
\eq{susy}, and solve for $A$ in terms of $\f$:

\be
e^A = \left( {2\f\over
1+\epsilon\f^2}\right)^{\mbox{${\epsilon\over 2 a^2}$}} \ ,
\la{ea}
\ee

where we have set a multiplicative integration constant equal to 2 for
convenience. Thus, the metric takes the form

\be
 ds^2 = \left( {2\f\over 1+\epsilon\f^2}\right)^{\epsilon/a^2}
(-dt^2+dx^2) + {1\over 16m^2a^4 } \left( {1+\epsilon\f^2\over
1-\epsilon\f^2}\right)^2  \left({\f'\over \f}\right)^2 dr^2 \la{m2}
\ee
\smallskip

It is straightforward to verify that all the field equations are
satisfied by this metric and the ansatz \eq{m}.

The fact that $\f$ is not determined by the equations of motion is a
consequence of having freedom in reparametrizing the radial coordinate
$r$. Indeed, the function $\f$ can be determined by performing a
$\f$-dependent $r$-coordinate transformation. A convenient such
transformation is

\be
 r \quad \rightarrow \quad {\tilde r}= M \left({1+\f^2(r)\over
1-\f^2(r)}\right)^2\ ,
\la{ct}
\ee

where $M$ is an integration constant. We next analyze the compact and
non-compact cases separately.


\subsection{The Case of $H^2$ ($\epsilon =-1$) }


The inversion of \eq{ct} yields the ${\tilde r}$-dependence of $\f$

\be
\f =\left({\sqrt{r}-\sqrt{M} \over
\sqrt{r}+\sqrt{M}}\right)^{1/2}\ .
\la{h1}
\ee

where the tilde on $r$ has been dropped for notational convenience. In
obtaining this result, we have chosen the positive root in \eq{ct}. The
negative root gives an expression for $\phi$ which diverges at $r=M$.
Note that $r\ge M$ implies $ \f \ge 0$ in accordance with the fact that
$\f$ is the stereographic coordinate of $H^2$.

The transformation \eq{ct} turns the metric \eq{m2} into

\be
ds^{2}= \left(\frac{r}{M} - 1 \right)^{\frac{-1}{2a^{2}}} (-dt^{2} +
dx^{2}) + \frac{1}{64 m^{2}a^{4} r^{2}} \left( \frac{r}{M} -1
\right)^{-2}dr^{2}.\label{ali}
\ee

This metric has no horizons and there is a naked singularity. To see
this, we first let $r\to r+M$ and then define a new radial coordinate
$ u = M/r$. The metric then takes the form

\be
ds^{2}= \left(u\right)^{\frac{1}{2a^2}}  (-dt^{2}+dx^{2})
+ \frac{1}{64 m^{2} a^{4} ( u + 1)^{2}} du^{2}.
\la{m5}
\ee

The asymptotic geometry near $u=\infty$ is $AdS_{3}$. The metric has a
singularity at $u=0$, while it is regular at other points.  The fact
that $u=0$ is a genuine singularity can be seen from the curvature
scalar associated with this metric, given by

\be
R={8 m^2 (u+1)\over u^2} \left[ 8a^2-3(u+1)\right]\ ,
\ee

which clearly diverges for $u=0$.  The implications of this naked
singularity for the cosmic censorship conjecture remains to be
investigated. 

Finally, we find that the $AdS$ energy per unit length for the string 
metric \eq{m5} vanishes. Actually, the commutator of two  $N=(2,0)$ 
supersymmetry transformations can be shown to vanish at radial 
infinity, but one cannot deduce from this alone that the $AdS$ 
energy vanishes. This is due to the fact that the result is 
a combination of the true Lorentz rotations and  translations in $AdS_3$. 
A more convenient method to pin down the $AdS$ energy for the case at hand is 
due to Hawking and Horowitz \cite{hawking}, and  applying this method, we 
indeed find the result stated above, namely  the vanishing of the $AdS$ 
energy for our solution.


\subsection{The Case of $S^2$ ($\e =1$) }


The inversion of \eq{ct} for the upper hemisphere $S^2_+$  yields

\be
\f_+ =\left({\sqrt{r}-\sqrt{M} \over
\sqrt{r}+\sqrt{M}}\right)^{1/2}\ .
\la{s1}
\ee

For the lower hemisphere $S^2_-$ we obtain

\be
\f_- =\left({\sqrt{r}+\sqrt{M} \over
\sqrt{r}-\sqrt{M}}\right)^{1/2}\ .
\la{s4}
\ee

Note that $r\ge M$ in both cases, in accordance with the fact that
$\f_\pm$ are the stereographic coordinates of $S^2$ such that $ 0\le
\f_+ \le 1$ and $1 \le \f_- < \infty$.  In fact, $\f_+$ and $\f_-$
constitute a well defined map from spacetime into $S^2$.\\

With the scalar field specified as in \eq{s1} or \eq{s4}, the metric (\ref{m2}) 
becomes

\be
ds^2 = \left(1-{M\over r}\right)^{1\over 2a^2} (-dt^2+dx^2)
+ {1\over 64m^2 a^4 r^2} \left(1-{M\over r}\right)^{-2} dr^2\ .
\la{m3}
\ee

Note that for this case $1/a^{2}$ is quantized to be an integer. This
metric is asymptotically Minkowskian. Moreover, there is a horizon at
$r=M$, and the near horizon geometry is $AdS_3$. The Hawking
temperature of this black string can be readily shown to be vanishing.
Thus, we expect this solution to be quantum mechanically stable.

The curvature scalar associated with the metric \eq{m3} is

\be
R=  {64 m^2 a^2 \over r^2} \left[ Mr - M^2\left(1+{3\over 8a^2}\right)
\right]
\ee

which is regular at $r=M$. This formula also shows that there is a
singularity at $r=0$. However, for some values of the parameter $a$
the singularity cannot be reached by the observers outside the
horizon.  To investigate this point, let us consider the geodesic
equation. Let $\xi^\m$ be tangent to an affinely parametrized
geodesic, and let us define the conserved quantities associated with
the two translations on the string worldsheet as $E=-\xi \cdot \del /
 \del t $, $P=\xi \cdot \del /\del x$. Then the geodesic equation
associated with the metric \eq{m3} takes the form

\be
{1\over 64 m^2 a^4} \left({{\dot r}\over r}\right)^2 = \a \left(1-{M\over
r}\right)^2 +(E^2-P^2)\left(1-{M\over r}\right)^{(4a^2-1)/2a^2}\ ,
\la{geo}
\ee

where the dot denotes derivative with respect to an affine parameter,
$\a=0$ for null geodesics and $\a=-1$ for timelike geodesics. For
$\a=-1$, the geodesics can not reach the horizon. Indeed, there is a
turning point corresponding to the vanishing of the right hand side of
\eq{geo}. For $\a=0$, a simple analysis of \eq{geo} near the horizon
shows that when $1/2a^2$ is an even integer the region $r<M$ is
accessible, but not accessible when $1/2a^2$ is an odd or half integer.
For the former case, we need to extend the definition of $\f$ to the
region $r <M$. However, Einstein equations imply that $C^2 =M/r$ and
thus $C>1$ for $r<M$. On the other hand, we see from its definition
that $ C^2 \le 1$ for any value of $\f$. Therefore, we can not extend
the solution to the region $r<M$ when $1/2a^2$ is an even integer.

To summarize, we have physically well defined black string solutions for

\be
{1\over a^2 } = 1,2,3\ {\rm mod}\ 4\ .
\ee

In these cases the timelike or null geodesics can not penetrate the
horizon located at $r=M$, and the field $\f$ need not be extended to
the region $r<M$.

For $a^2=1/2$, the metric \eq{m3} coincides with the metric found by
Horne and Horowitz \cite{hh} obtained from a different starting point,
namely low energy limit of a string theory in $2+1$ dimensions
described by the Lagrangian

\be
e^{-1}{\cal L} = e^{\f} \left(
R + \del_\m \f\, \del^\m \f -\ft1{12}
H_{\m\n\r} H^{\m\n\r}+ {8\over k} \right)\ ,
\la{hh}
\ee

where $H=dB$ and $k$ is a constant. In Einstein frame, this Lagrangian
takes the form

\be
e^{-1}{\cal L}= R-\del_\m\f \del^\m\f-\ft12 e^{4\f}H^2 + \ft8{k}
e^{-2\f} \ .
\la{ef}
\ee

The metric \eq{m2} is a solution for this theory, with the dilaton
given by $\f = {\rm ln}\,(r\sqrt{k/2}\,) $. What we have shown here is
that not only this metric is a solution of two rather different
theories but it is also supersymmetric. We note that the string theory
which should generate our matter coupled $N=(2,0)$ $AdS_3$ supergravity
model remains to be determined.

Finally,  we note that the mass per unit length for our general string
solutions  in the $S^2$ sigma model case  can be conveniently deduced from
the algebra  of supercharges, since these solutions are asymptotically
Minkowskian.  A standard procedure which makes use of the Nester two-form 
(see, for example, \cite{ghw,less}) yields the
result

\bea
[Q_{\ve_1}, Q_{\ve_2}] &=& \lim_{r \rightarrow \infty}
{\e^{012}\over \sqrt {-g}} {\bar\varepsilon}_1 
\left(D_2 +m\c_2 C^2\right)\ve_2
\nn\w2
 &=& {\bar\varepsilon}_1 \left( \c^0 P_0 + \c^2 P_2\right)\ve_2\ ,
\eea

where $P_0=P_2= 8m^2a^2 M$, and $0,1,2$ refer to the time, radial and 
$x$-directions, respectively,  in coordinate basis ($\e^{012}=1$, 
in our conventions). Thus, the string has  mass and momentum per unit length 
$ 8m^2a^2 M$. We refer the reader to  \cite{h2} for a  study of various
aspects of this phenomenon.


\section{The $N=(1,0)$ AdS$_3$ Truncation}


The $N=(2,0)$ model described in the previous sections admits a
truncation to $N=(1,0)$ supersymmetric $AdS_3$ supergravity coupled to
$n$ scalar multiplets. It is straightforward to check that the
following truncation is consistent:

\bea
&& A_\m=0\ ,\qq P_\m^{1r}=0\ ,
\w2
&& (1-\C_2) \psi_\m=0\ , \qq (1-\C_2) \l^r=0\ , \qq
(1-\C_2) \ve =0\ .
\nn
\eea

The condition $P_\m^{1r}=0$ amounts to setting $n$ of the original $2n$
scalar fields equal to zero. A convenient way to realize this condition
is to parametrize $L$ as follows

\be
L=\left(\matrix{ (I-\e \f \f^T)^{1/2} &0& \f\cr 0&1&0\cr
 -\e \f^T &0& (1-\e \f^T\f)^2 \cr}\right)\ ,
\la{nr}
\ee \\

where $\f$ is an $n$-component column vector representing the $n$ real
scalars. From the definitions \eq{cf} and the representation
$(T_{ij})_I{}^J= (\,\e\, \d_{Ii}\,\d_{j}^J - i \leftrightarrow j\,) $,
it follows that

\be
C^{rs}=0\ ,\qq S^{2r}=0\ , \qq Q_\m =0\ ,
\ee

and

\bea
C &=& \left(1-\e \f^2\right)^{1/2}\ ,
\nn\w2
S^r &=& \f^r\ ,
\nn\w2
P_\m^r &=&
\e\left[ (1-\e\f\f^T)^{-1/2}\del_\m \f\right]^r \ ,
\nn\w2
Q_\m^{rs} &=& 2\f^{-2}\left[1-(1-\e\f^2)^{1/2}\right]\,
\f^{[r}\del_\m\f^{s]}\ ,
\eea
\smallskip

where $\f^2 \equiv \f^T\f$, $S^r\equiv S^{1r}$ and $P_\m^r \equiv
P_\m^{2r}$. The identities \eq{lemmas} now take the form

\be
\del_\m C= -P_\m^r S^r\ , \qq D_\m S^r= \e P_\m^r C\ .
\la{lem2}
\ee

Performing the truncation procedure described above, we are left with
the $N=(1,0)$ $AdS_3$ supergravity multiplet consisting of a dreibein
and a single Majorana gravitino, and $n$ copies of $N=(1,0)$ scalar
multiplets each one of which contain a real scalar and a Majorana
spinor. The generic manifolds parametrized by the scalar fields are now

\be
N_+= {SO(n+1)\over SO(n)}\ , \quad\quad N_-=
{SO(n,1)\over SO(n)}\ .
\la{pm2}
\ee

The truncation of the Lagrangian \eq{a} gives

\bea
e^{-1}{\cal L} &=&  {1\over 4} R
+ {1\over 2} \e^{\m\n\r}\pb_\m D_\n \psi_\r
-{1\over 4a^2} P_\m^r P^\m_r
\nn\w2
&& +{1\over 2}\,\lb_r\c^\m D_\m \l^r
   + {1\over 2a}\, \lb_r \c^\m\c^\n \psi_\m P_\n^r
   -{m\over 2}\, \pb_\m\c^{\m\n}\psi_\n C^2
\nn\w2
&&  -2 ma\,\pb_\m \c^\m\l_r C S^r
    -{1\over 2} m (1\pm 4 a^2)\, \lb^r\l_r C^2
\nn\w2
&& + {2 ma^2}\,\lb_r \l_s S^r S^s
+{2 m^2} C^2(C^2 -2 a^2 S^rS^r)\ ,
\la{atr}
\eea

which has the local $N=(1,0)$ supersymmetry

\bea
\d e_\m{}^a &=& -\eb\c^a\psi_\m\ ,
\nn\w2
\d \psi_\m &=& D_\m \ve +m \c_\m C^2 \ve\ ,
\nn\w2
L^I \d L_I{}^r &=& a\, \eb\, \l^r\ ,
\nn\w2
\d\l^r &=& \left( -{1\over 2a} \c^\m P_\m^r
- 2 ma C S^r\right)\ve\ .
\la{str}
\eea

The index $I$ now labels the $n+1$ dimensional representation of
$SO(n+1)$ or $SO(n,1)$, and the matrices $(L^I,L_r^I)$ together form an
element of these groups. The latter can be represented by \eq{nr} with
the $(n+1)$'st row and column deleted.

Clearly the black string solution of the $N=(2,0)$ model described in
Section 4 is also a solution of the $N=(1,0)$ model given here.\\

The $C$- and $S$-functions defined in \eq{lemmas} arose as a
consequence of the commutator \eq{com}. It is noteworthy that these
functions still arise in the $N=(1,0)$ model despite the fact that the
commutator \eq{com} no longer occurs. Indeed, as far as supersymmetry
is concerned, all that is required of the $C$ and $S$-functions is that
they obey the relations \eq{lem2}. This suggests the possibility of a
more general solution for them. To see this, let us consider the case
of $SO(1,1)$ scalar manifold. In that case the $C$- and $S$-functions
take the simple form

\be
C={\rm cosh}\,\vf\ ,\qq S={\rm sinh}\,\vf \ ,
\la{cs5}
\ee

where we have defined $ \phi = {\rm sinh}\,\vf$. The bosonic Lagrangian
then becomes

\be
e^{-1} {\cal L}= \ft14 R -{1\over 4a^2} \del_\m\vf \del^\m \vf -
V(\vf)\ ,
\ee

where

\be
V=2m^2 {\rm cosh}^2\,\vf \left( 2a^2{\rm sinh}^2\,\vf-{\rm cosh
}^2\,\vf \right)\ .
\la{v1}
\ee

A more general solution of the defining relation \eq{lem2} is

\bea
C = a_1 e^{\vf} + a_2 e^{-\vf}\ ,\qq
S= -a_1 e^{\vf} + a_2 e^{-\vf}\ ,
\eea

where $a_1,a_2$ are arbitrary real constants. These functions define a
family of $N=(1,0)$ $AdS_3$ coupled to a single scalar multiplet, with
potential

\be
V(\vf)= 4 m^2 a^2 \left[ \left(a_1 e^{\vf} + a_2
e^{-\vf}\right)^4-{1\over 2a^2} \left(a_1^2 e^{2\vf} -a_2^2
e^{-2\vf}\right)^2 \right]\ .
\la{pot}
\ee

In fact, $a_1/a_2$ is the only independent parameter, due to the
freedom in rescaling the parameter $m$. For $a_1/a_2=1$, one obtains
the $N=(1,0)$ truncation of the $N=(2,0)$ model discussed above.


\section{Conclusions}


We have constructed the coupling of $n$-complex dimensional
Kahler sigma models of certain type to the $AdS_3$ supergravity with
$N=(2,0)$ supersymmetry and we have obtained the black string solutions
of this model. We have also obtained the $N=(1,0)$ truncation of our
model, which still admits a potential as well as the solutions of the
$N=(2,0)$ model discussed here. 

Our models generically depend on two parameters which characterize the sizes of 
the $AdS_3$ and the  sigma model manifold, respectively.Moreover, the
geometry of  the sigma models can  be compact or non-compact. The
properties of the string  
solutions presented  here depend on the geometry of the sigma model. In
the compact  case, we have found 
asymptotically Minkowskian  black string solutions, while in the non-compact 
case we have found asymptotically $AdS_3$ string solutions with naked 
singularities. In the  former case, our solution is found to coincide with 
that of Horne and  Horowitz \cite{hh} for a particular radius of the 
compact sigma model manifold.

A previously constructed \cite{it} coupling of a Kahler sigma model to
$N=(2,0), AdS_3$ supergravity differs from our model significantly in
that the scalar fields of that model are neutral under the $R$-symmetry
group $U(1)$. One consequence of having assigned a $U(1)$ charge to the
scalar fields is the emergence of a potential, which plays a
significant role in the determination of our black string solutions.

As for the classical solutions of our model, it is natural to seek
supersymmetric black holes. Indeed, we have searched for solutions of
the form $ds^2=- e^{2A}dt^2+e^{2B}dr^2 +r^2 d\vf^2$. Setting the scalar
field equal to zero reduces the equations of motion to those of pure
anti de Sitter supergravity which is known to have the BTZ black hole
solution \cite{it}. However, if we insist on non-vanishing scalar
fields, then the supersymmetry condition, together with the field
equations, leads to a solution which upon coordinate transformations
can be brought to the string solution of the form \eq{m3}.

It would be interesting to find a solution in which the Maxwell field
plays a role. In this context, we note that the Einstein's equation
rule out nonvanishing gauge fields if we take all the fields to be only 
$r$-dependent. Presumably, therefore, one should allow $x$ dependence as 
well.

A natural extension of our model would be the introduction of higher
than $N=(2,0)$ supersymmetry. An interesting case to consider is the
matter coupled $N=(4,4)$, $AdS_3$ supergravity model which arises from
the compactification of the $D=6$, $N=(2,0)$ supergravity coupled to
$n$ tensor multiplets, which has its origin in the $K_3$ compactification of 
Type IIB string. 

The question of how the models we have presented in this paper can be 
obtained from any compactification of M-theory or, for that matter, 
any higher dimensional supergravity theory remains open. The exact form of
the CFT dual of our model formulated on the boundary of $AdS_3$ also
remains to be found.

Finally, we note that the model constructed in this paper does not
involve any two-form potentials. These potentials would not
describe propagating degrees of freedom but they might be useful 
in producing the low energy supergravity theory in the bulk in a form which 
is more appropriate in the string theory or the boundary CFT context. In the 
models we  have constructed here, it is natural to introduce $(n+2)$
two-form 
potentials which  form an $SO(n,2)$ vector. The form of the scalar
potential in the 
action is then expected to  change, but elimination of the two-form potential 
from the action through  its equations of motion should yield the scalar
potential  of the model presented here.

\bigskip \bigskip \noindent{\large \bf Acknowledments}

\medskip

We wish to thank R. Argurio, J. de Boer, M.J. Duff, M. Gunaydin, P.S.
Howe, J.X. Liu, J. Maldacena, K.Skenderis, P.K. Townsend and V. Zhukov
for fruitful discussions and I. Rudychev for his collaboration at early
stages of this work. This research has been supported in part by NSF
Grant PHY-9722090.

\end{document}